\documentclass[]{article}
\usepackage{amsmath,amssymb,float,geometry,pslatex}
\usepackage{graphicx}
\usepackage{mathptmx} 
\usepackage{epstopdf}
\usepackage{authblk}
\usepackage[numbers, sort, comma, square]{natbib}
\def\bes{\begin{eqnarray}}\def\ees{\end{eqnarray}}\def\beq{\begin{equation}}\def\eeq{\end{equation}}

  \def\ba#1\ea{\begin{align}#1\end{align}}
\def\bsa#1#2\esa{\begin{subequations}\label{#1} \begin{align}#2\end{align} \end{subequations}}

\def\O{{\mathcal{O}}}
\allowdisplaybreaks

\def\O{\mathcal{O}}

\def\bes{\begin{eqnarray}}\def\ees{\end{eqnarray}}\def\beq{\begin{equation}}\def\eeq{\end{equation}}

\def\ba#1\ea{\begin{align}#1\end{align}}

\def\bsa#1#2\esa{\begin{subequations}\label{#1}
\begin{align}#2\end{align} \end{subequations}}

\def\f{\frac}

\begin{document}

\title{Rogue Wave Morphology in Broadband Nonbreaking Seas}

\author[1]{Q. Guo}
\author[2]{L.-A. Couston}
\author[1]{M.-R. Alam\thanks{reza.alam@berkeley.edu}}

\affil[1]{Department of Mechanical Engineering, University of California Berkeley, CA 94720}
\affil[2]{CNRS, Aix Marseille Univ, Centrale Marseille, IRPHE, Marseille, France}
\date{}
\maketitle

\begin{abstract}

Here, we show through a statistical analysis that rogue waves in broadband non-breaking seas are spatially asymmetric. In addition to the top-down asymmetry due to nonlinear effects, we show that the two troughs adjacent to the rogue wave crest are generally of different depths, which is unlike the conventional picture of rogue waves with symmetric fore and aft troughs often obtained from model equations. The rogue-wave trough asymmetry is demonstrated for sea states 4 to 6 on Douglas Sea Scale. Considering the deepest trough leads to approximately 10\% increase in the calculation of the mean rogue-wave height compared to previous results for rogue waves reported with symmetric troughs. This implies that estimates of rogue-wave trough-crest amplitudes based on model equations should be re-assessed upward for all realistic oceanic conditions. 

\end{abstract}

\section{Introduction}
Rogue waves, also known as freak waves or monster waves, are unexpectedly large-amplitude waves in the ocean. Rogue waves are usually defined as waves with peak to trough height larger than two times the significant wave height, which, in relatively common rough sea state 6 on Douglas Sea Scale, correspond to 25m or larger waves. Observations of extreme waves by the oil and shipping industry across the world's oceans \cite{Christou2014} have revealed the unexpectedness of rogue waves: rogue waves have a much higher rate of occurrence than predicted by classical sea state spectra without the information of phases \cite{Janssen2003}, and thus defy Gaussian statistics. Our understanding of the unexpected nature of rogue waves has yet significantly improved over the past few decades, and the high occurrence rate can now be explained in part by nonlinear mechanisms, such as the Benjamin-Feir instability \citep[see e.g.][]{Dysthe2008}. 

In addition to the occurrence rate of rogue waves and to the predictability horizon \cite{Alam2014}, which provides information on how much in advance and how accurately we can predict the formation of rogue waves for a given sea state, the morphology of rogue waves is of fundamental importance and of growing interest in maritime design \cite{GuedesSoares2005,Soares2008}. A popular theoretical model for rogue wave dynamics is the nonlinear Schr\"odinger equation (NLS), for which analytical solution exist. Self-focusing solutions of NLS that can potentially represent rogue waves include the first-order Peregrine Breathers \cite{Peregrine1983}, and the higher-order rogue wave triplets \cite{chabchoub2013}, triangular and polygon wave patterns \cite{He2012}, and circular rogue wave clusters \cite{kedziora2011}. The averaged rogue wave profile in the framework of higher-order NLS-type equations gives similar symmetric shape\cite{Socquet-Juglard2005}. First-order and higher-order NLS solutions always exhibit a high level of symmetry in the horizontal-temporal $(x,t)$ plane with respect to the main crest, hence suggesting that rogue waves have symmetric fore and aft troughs, a characteristic shared by averaged rogue wave profiles from higher-order NLS-type equations \cite{Socquet-Juglard2005}. NLS-type equations are, however, limited to narrow-banded wave spectra. Therefore, a natural and important question can be formulated: should we expect rogue waves predicted by general wave models, i.e. valid for broadband spectra, to have symmetric troughs?  

For a broadband wave spectrum, existing results on the morphology of rogue waves  \citep[e.g.][]{xiao2013} have suggested that the troughs next to the main crest have similar shape and depths. The auto-correlation function of sea surface in Gaussian sea states is also shown to exhibit symmetric averaged profile of rogue waves\cite{Phillips1992}. However, observations of rogue waves in the oceans show a different picture. For instance, while the \emph{New Year Wave} measured near the Draupner platform \cite{Haver2004} has preceding and succeeding troughs of almost the same depth, the Andrea wave measured in a storm crossing the North Sea \cite{Magnusson2012} has troughs of substantially different depths. Furthermore, the rogue wave measured in the North Cormorant field \cite{GuedesSoares2005} shows a much deeper trough prior to the main crest compared to the one trailing behind \citep[see also][for other observations of rogue waves]{Sand1990,GuedesSoares2003,Mori2002,Garett2009}. Differences in trough depths have also been reported for the case of broadband long-crested seas (i.e. amenable to two-dimensional studies), as shown by wave basin experiments \cite{Klein2011}, few numerical simulations \cite{Touboul2006}, and field measurements \cite{Gemmrich2017a}.  

Here we demonstrate that rogue waves are asymmetric with respect to the main crest for broadband nonbreaking seas, which can lead to underestimation of rogue wave trough-to-crest height.  This strong asymmetry (referred to as trough asymmetry) of rogue waves emerging from broadband sea states has not been obtained by previous averaging methods based on proper orthogonal diagonization (POD) \cite{xiao2013} and auto-correlation function \cite{Phillips1992}. We will demonstrate that the trough asymmetry is more pronounced in space domain than in time domain, hence is difficult to capture with fixed buoy measurements. This may explain why the trough asymmetry has not yet been reported as a recurring feature of rogue waves in the field, since field observations are more often based on time series of wave elevation at fixed locations with buoys than on spatial profiles based on e.g. Synthetic Aperture Radar (SAR) images. Nevertheless, the reproduction of the New Year Wave in a wavetank that did yield a rogue wave with strong trough asymmetry \citep[cf. second rogue-wave occurrence in figure 3 of][]{Clauss2009} and a SAR image-based reconstruction of a rogue-wave spatial profile with the right trough more than three times deeper than the left trough \citep[cf. figure 6 in][]{Muller2005} are evidence supporting the existence of trough asymmetry for rogue waves in the oceans. 

\section{Method}
In order to model the ocean surface dynamics, here we consider the two-dimensional potential flow equations in Zakharov form \cite{Zakharov1968}, i.e.
\bsa{eq:1}
\eta_t&=\phi^{S}_{z}(1+\eta_x^2)-\phi^{S}_{x}\eta_x,  &\text{at } z=\eta(x,t),\label{eq:1a}\\
\phi^{S}_{t}&=-g\eta-1/2(\phi^{S}_{x})^2-1/2\phi^{2}_{z}(1+\eta_{x}^2), &\text{at } z=\eta(x,t), \label{eq:1b}
\esa
derived from the Navier-Stokes equations under the assumption of an inviscid, irrotational, incompressible and homogeneous fluid. The Cartesian coordinates system is aligned with the mean free surface with $z$ axis vertical upward and $x$ axis horizontal, $\eta$ is the free surface elevation and $\phi^{S}(x,t)=\phi(x,z=\eta,t)$ the velocity field at the free surface. We solve equations \eqref{eq:1} with a phase resolved High-Order Spectral method (HOS). We assume that $\phi$ can be expressed in perturbation series as $\phi=\sum_{m=1}^{M}\phi^{(m)}$ with $\phi^{(m)}\sim\O(\epsilon^m)$, where $\epsilon\ll 1$ is the wave steepness ($\eta$ expanded similarly), such that, regrouping terms of same order results in a system of $M$ forced linear partial differential equations. The equations are then efficiently solved sequentially  using a spectral decomposition of $\phi_S$ and $\eta$ in $N$ Fourier modes. The results converge exponentially fast with $N$ and $M$ up to wave steepness $\epsilon\approx 0.35$ \cite{Dommermuth1987}, allowing to capture high-order nonlinearities, which are essential to the rogue wave dynamics, at a relatively low computational cost (we typically use $N\sim \O(1000)$, $M\sim\O(10)$).

 \begin{table}
 \caption{Sea states considered here with relevant physical  and simulation parameters, along with key results regarding rogue wave asymmetry. The peak wavenumber $k_p$ is obtained from $T_p$ using the linear dispersion relation for surface waves in deep water. RWs is the number of rogues waves obtained from $\O(1000)$ simulations and Prec gives the number of times the deepest trough precedes the rogue wave crest such that the profile is flipped when averaging according to equation \eqref{eq:4}. $\bar{\eta}_D/\bar{\eta}_S$ is the mean ratio of the deepest to shallowest trough, and $\bar{H_r}=\bar{\eta}_C+\bar{\eta}_D$ is the deepest trough to crest height of averaged rogue wave profile. \vspace{0.2cm}}
 
 \centering
 \begin{tabular}{ccccccccccccccc}
 \hline
  \begin{tabular}{@{}c@{}}Sea \\ State\end{tabular} & $H_s$ (m) & $T_p$ (s) & $\epsilon_p$ & $h$ (m) & $k_ph$ & $M$ & $\frac{\delta x}{\lambda_p}$ & $\frac{\delta t}{T_p}$ & RWs & Prec & $\frac{\bar{\eta}_D}{\bar{\eta}_S}$ & $ \bar{H_r}$ \\
 \hline
   4  & 1.875  	& 8.8 	& $\frac{5}{10^3}$ & 300 & 15.6 & 4  & $\frac{18}{10^3}$ & $\frac{1}{128}$ & 48 & 26 & 2.04 & 2.09\\
   5  & 3.25  	& 9.7 	& $\frac{7.8}{10^3}$ & 300 & 12.8 & 1-5 & $\frac{15}{10^3}$ & $\frac{1}{128}$ & 72 & 39 & 1.95 & 2.28\\
   6  & 5     	& 12.4 	& $\frac{9.4}{10^3}$ & 300 & 7.9 & 4 & $\frac{9}{10^3}$ & $\frac{1}{128}$ & 57 & 30 & 2.02 & 2.14\\
 \hline
 \end{tabular}\label{toutou}
 \end{table}

Table \ref{toutou} summarizes the different sea states (4, 5, 6 on Douglas scale) considered as well as key physical and numerical parameters. The initial free-surface at $t=0$ is given by
\ba{}\label{init}
\eta = \eta^{(1)} = \sum_{n=1}^N \sqrt{2S_{dis}(k_n)\delta k_n} \exp^{i(k_n x+\theta_n)},
\ea
where $S_{dis}(k)$ is a discretized version of the JONSWAP spectral density respect to wave number. $S(k)$ is related with JONSWAP spectral density function respect to angular frequency $S(\omega)$ as $S(k)=S(\omega)C_g$, where $C_g$ is the group velocity and $S(\omega)=\frac{\alpha H_s^2\omega_p^4}{\omega^5}\exp(\beta)\gamma^{\delta}$ \cite{Carter1982}. In equation \eqref{init}, $\omega$ is the wave radial frequency, $\omega_p=2\pi/T_p$ is the peak wave radial frequency, and $H_s$ is the significant wave height, i.e. the mean trough-to-crest wave height of the one third highest waves (see appendix \ref{a1} for additional details); note that both $T_p$ and $H_s$ change with the sea state. The initial phase distribution $\theta_n\in(0,2\pi)$ in \eqref{init} is random with uniform distribution, allowing for different initial conditions for each sea state considered. Note that in order to avoid spurious modes known to contaminate numerical solutions \cite{Wu2004}, we introduce nonlinear effects gradually, i.e. we multiply nonlinear terms by $\hat{W}$ where $\hat{W}$ increases from 0 to 1 in $5T_{p}$. 

For each sea state, we run $\O$(1000) simulations so that $\O$(100) rogue waves are obtained under the criterion $H_{r}(t)>2H_{s}(t)$, where $H_{r}(t)$ is the maximum peak to adjacent trough height and $H_s(t)=4\sigma_\eta(t)$ is the significant wave height. With $\O$(100) rogue waves, the standard error of the mean rogue-wave profile is maximum (2\%) at the peak, hence the averaged profile is statistically converged. In the simulations, we search for rogue waves in the time window $100T_{p}<t<130T_{p}$ in order to allow high-order nonlinearities to develop (nonlinear effects beyond second-order develop in $t\sim\O(1/\epsilon^2)\approx 45 $ for $\epsilon \leq 0.15$). 

%


%
The averaged and normalized rogue wave profile obtained from our numerical simulations match well with the ones from field measured rogue waves analyzed in \cite{Christou2014}. Comparing the time-average profile of rogue waves from our simulations in sea state 5 with the time-average profile of more than three thousand rogue waves observed in multiple locations worldwide, we find that the maximum discrepancy is less than 6\% and occurs close to the rogue wave crest. The simulation and field averages lie within the averaged profile plus/minus the standard deviation of each other,  demonstrating that the rogue waves database generated numerically constitutes an appropriate representation of rogue-wave profiles in real seas.


\section{Results}\label{sec:3}

\begin{figure}
\centering
\includegraphics[width=2.55in,clip=false]{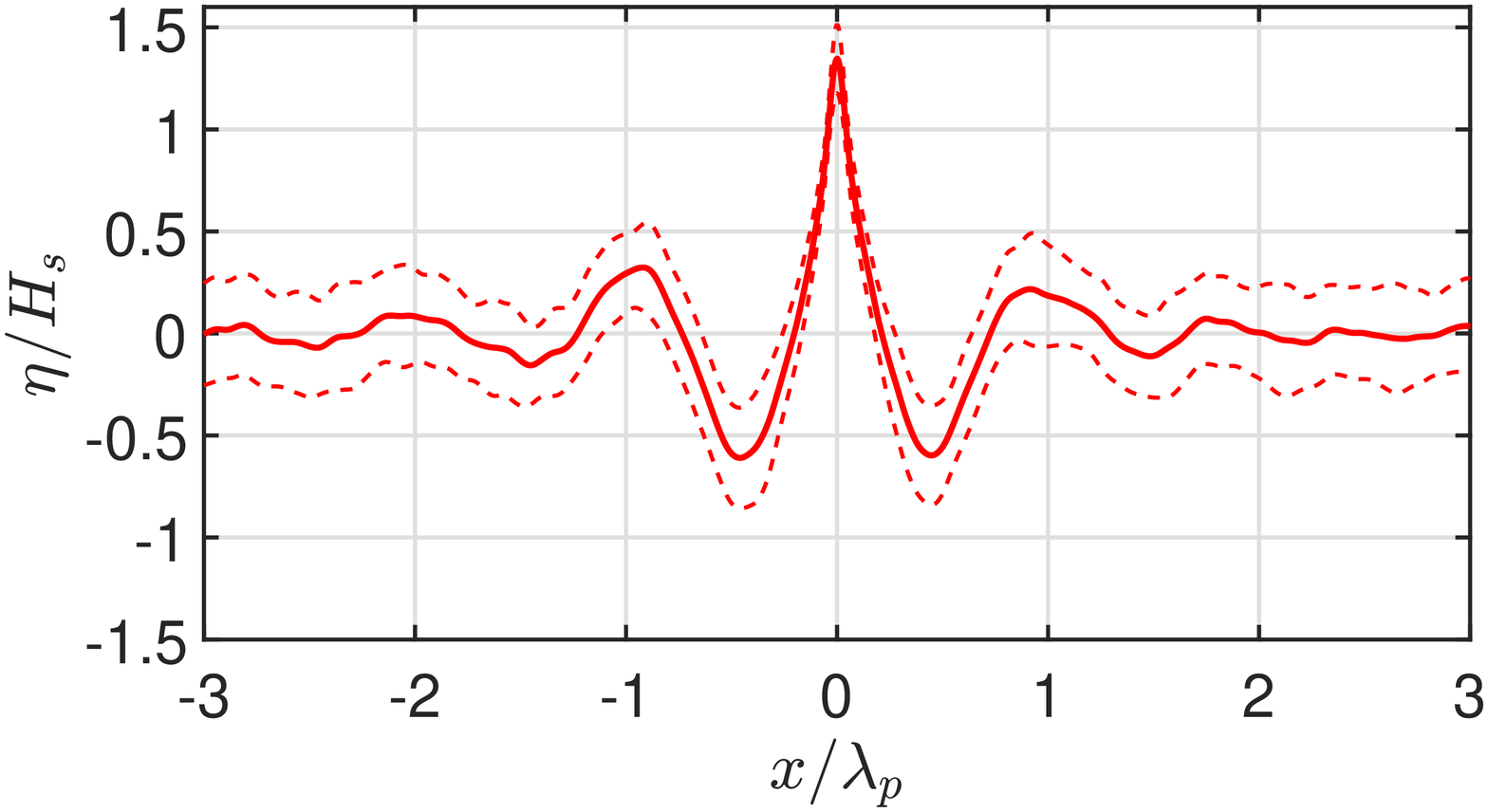}
\put(-155,90){(a)}
\includegraphics[width=2.55in,clip=false]{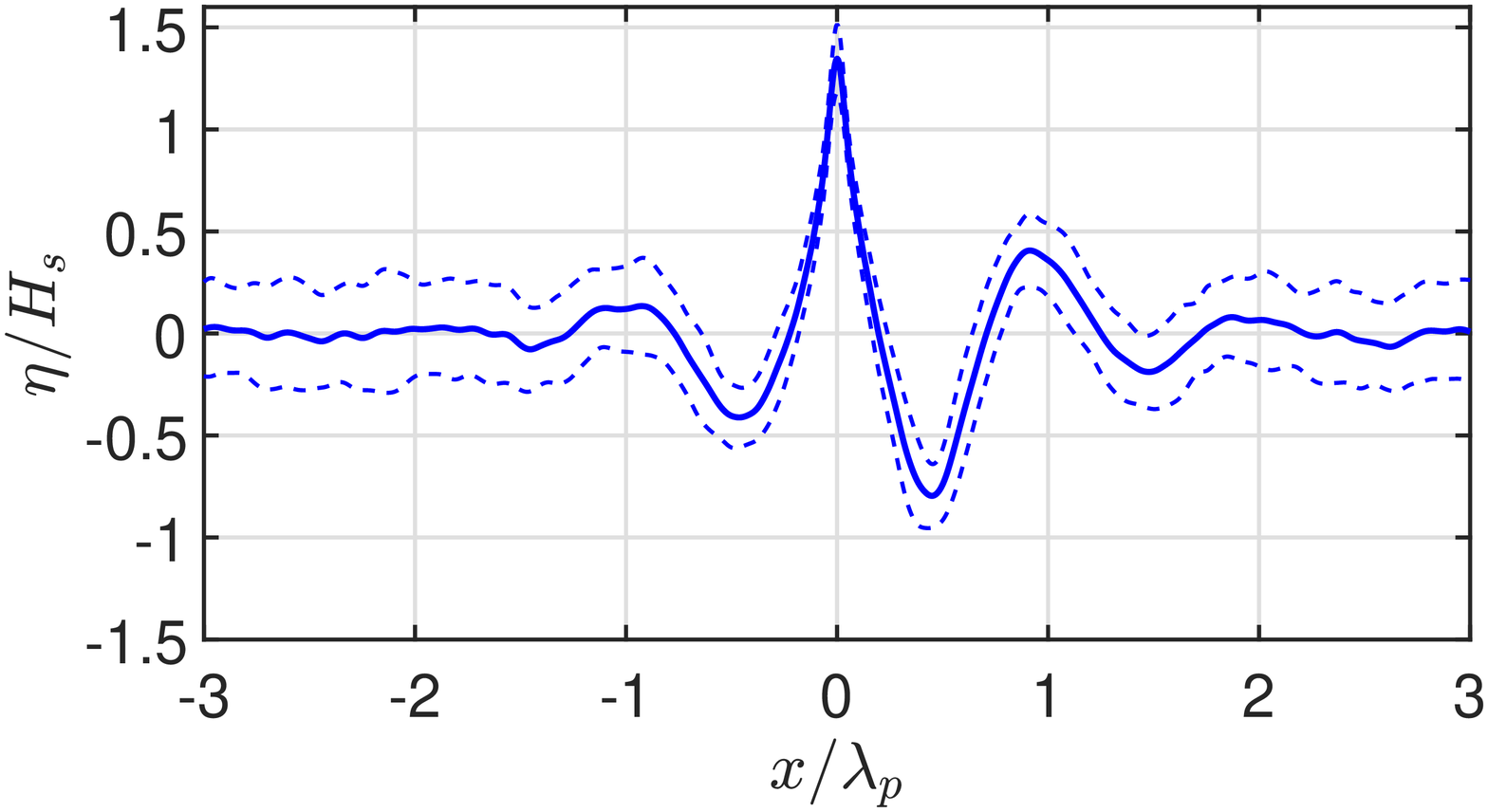}
\put(-155,90){(b)}\\
\vspace{-0.4cm}
\includegraphics[width=2.55in,clip=false]{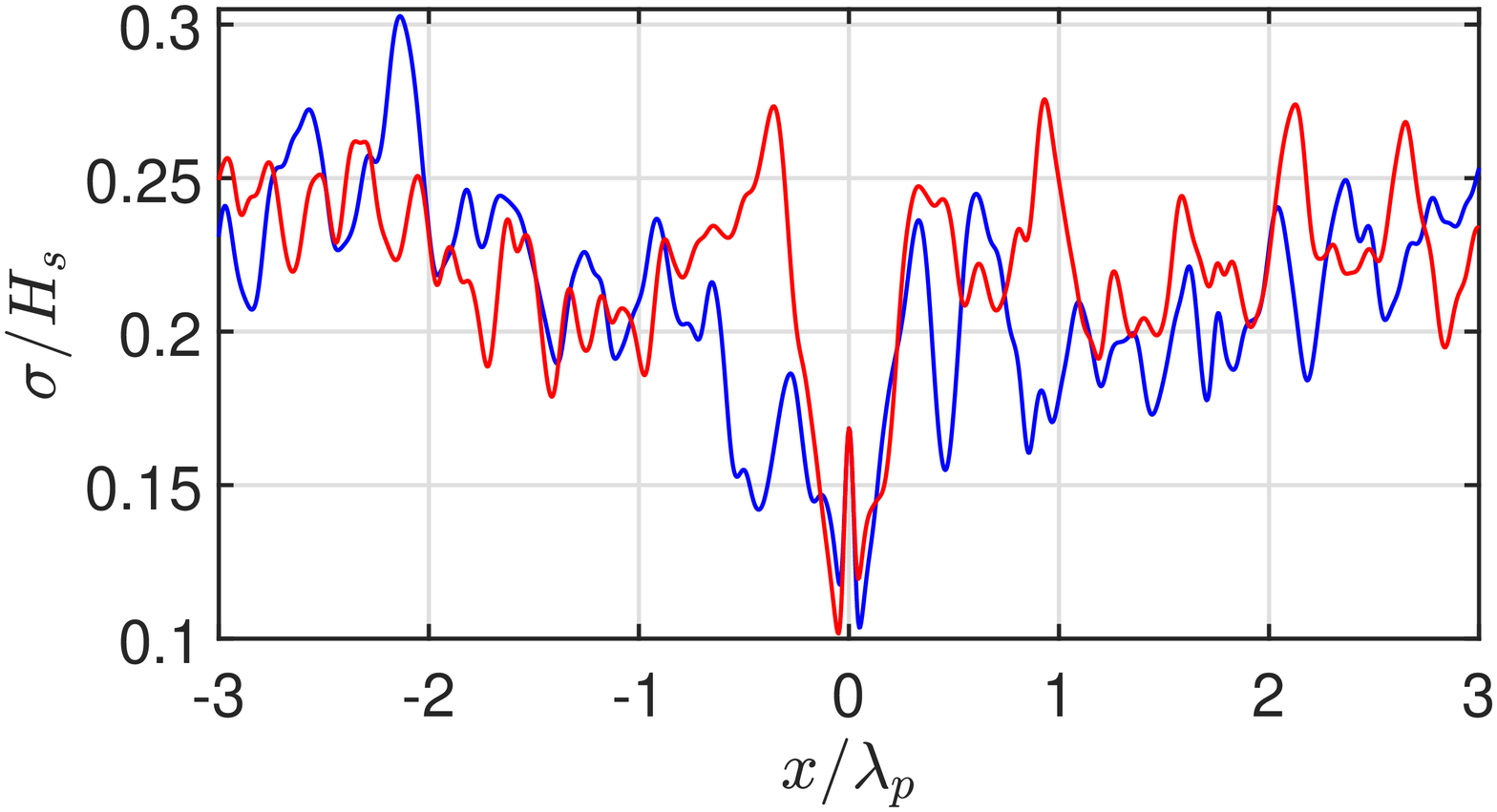}
\put(-155,90){(c)}
\includegraphics[width=2.55in,clip=false]{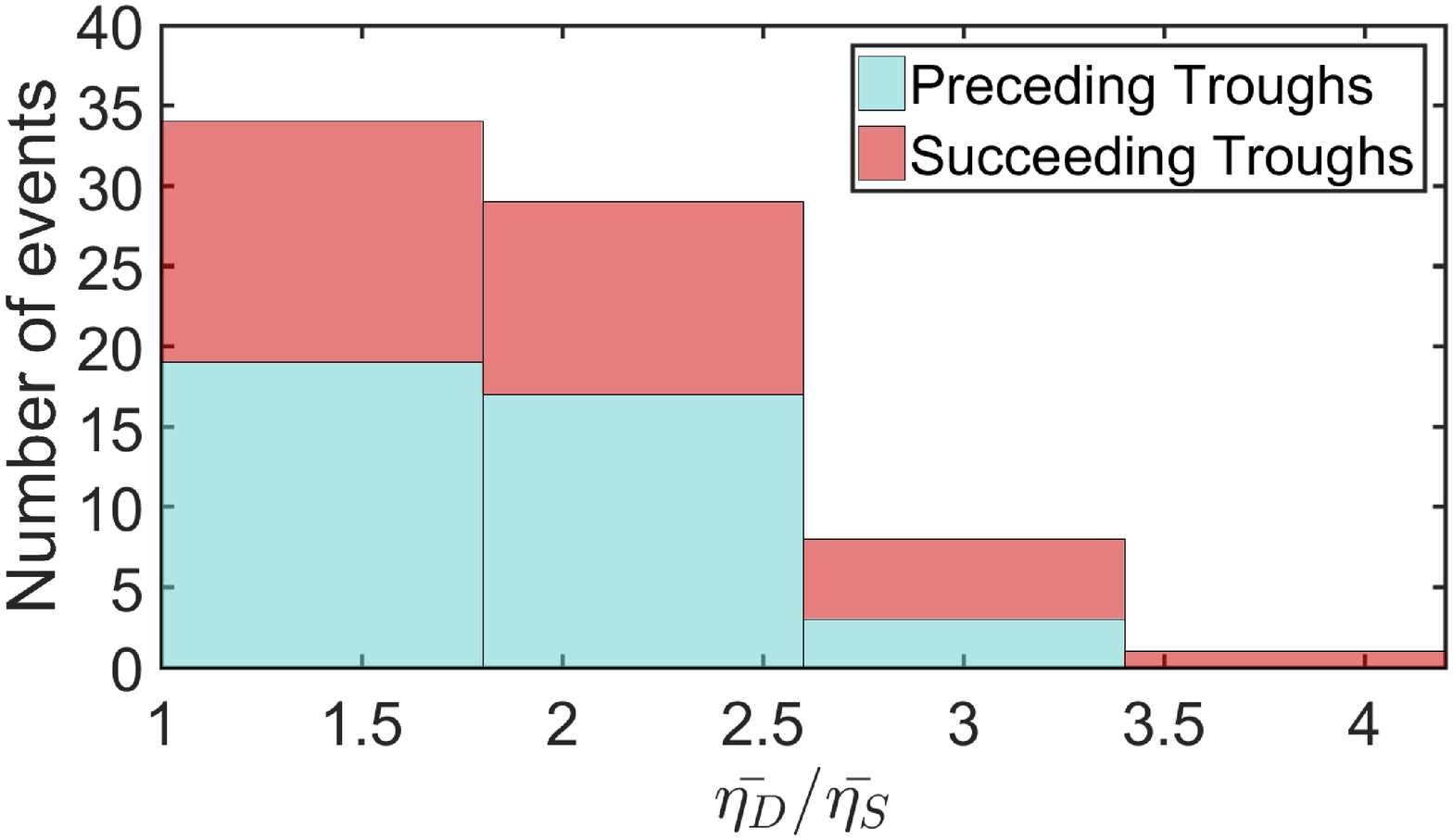}
\put(-155,90){(d)}\\
\vspace{-0.5cm}
\caption{Are rogue waves spatially symmetric? From the 72 rogues waves obtained for sea state 5 (see table \ref{toutou}), we compute the mean rogue wave profile $\bar{\eta}$ based on (a) equation \eqref{eq:3} (method i-), and (b) equation \eqref{eq:4} (method ii-). Method i- is commonly used in rogue wave research \citep[e.g.][]{xiao2013} and results in fore and aft troughs that are symmetric of each other with respect to the rogue wave crest. Method ii-, which involves flipping the troughs so as to keep the shallow troughs on the left hand-side of the crest, however, reveals the trough asymmetry of rogue waves. The standard deviation for each case is shown in figures (a)-(b) on top of the mean rogue wave profile (dashed lines) and in (c). The distribution of deep trough depth to shallow trough depth $\bar{\eta}_D/\bar{\eta}_S$ is shown in the form of two superposed histograms in (d). The bars with blue or red color are the histograms of rogue waves with the deep trough preceding or succeeding the main crest.}\label{fig:2}
\end{figure}

The main result of the paper is the average rogue wave profile $\bar{\eta}$ obtained from two different methods and shown in figures \ref{fig:2}a-b for sea state 5. The first method (method i-) is based on direct averaging of the crest-centered rogue wave profiles normalized by the instantaneous significant wave height $H_s(t)$, i.e.
\ba
\bar{\eta} = \frac{1}{\mathcal{R}}\sum_r \f{\eta_r(\f{x-x_r}{\lambda_p})}{H_s(t)} \label{eq:3}
\ea
at the time of rogue wave occurrence ($\mathcal{R}$ the total number of rogue waves in database, $x_r$ the location of rogue wave crest, $\lambda_p(t)$ the instantaneous peak wavelength), and results in the mean rogue wave profile shown in figure \ref{fig:2}a. The second method (method ii-) consists in averaging the rogue wave profiles with flipping, i.e. calculating
\ba
\bar{\eta} = \frac{1}{\mathcal{R}}\sum_r \f{\eta_r(p_r\f{x-x_r}{\lambda_p})}{H_s(t)}\label{eq:4}
\ea
where $p_r=1$ (resp. -1) when the deepest trough precedes (resp. succeeds) the highest crest, and results in the mean rogue wave profile shown in figure \ref{fig:2}b. The key point is that method -ii preserves the asymmetry of the trough depths surrounding the main crest, as can be seen from figure \ref{fig:2}b, whereas method -i, which is often used in rogue wave research \citep[e.g.][]{xiao2013}, loses this information (see figure \ref{fig:2}a). The standard deviation of the rogue wave profiles is plotted in figure \ref{fig:2}c for each approach. The standard deviation is high close to both trough locations when using method i-, suggesting that the trough depths distribution is widespread on both sides of the main crest. The standard deviation using method ii- is 20\% and 50\% less than that of method i- at succeeding and preceding trough, respectively. Figure \ref{fig:2}d shows the distribution of the deep-to-shallow trough depth ratio in the form of two superposed histograms; the histogram filled with blue (resp. red) color represents the ratio for deep troughs preceding (resp. succeeding) the main crest. Because the two histograms record about the same number of events, the occurrence rate of a deep fore trough is as high as of a deep aft trough. This explains why method i- loses the information on the rogue wave asymmetry: the deeper trough can occur on either side of the rogue wave, such that direct averaging gives the same mean fore and aft trough depth. We find that more than half the rogue waves have a deep trough more than twice as deep as the shallow trough. On average, the deep-to-shallow trough depth ratio is $\bar{\eta}_D/\bar{\eta}_S=1.95$, and the deepest trough-to-crest height $\bar{H}_r=\bar{\eta}_C+\bar{\eta}_D$ calculated from method ii- is 2.14, which, compared to 1.96 for method i-, is approximately 10\% larger (see table \ref{toutou}). It is important to note that, although not shown, normalized spatial profiles calculated using method ii- look similar qualitatively and quantitatively for all sea states (i.e. 4, 5 and 6). In particular, the normalized crest-to-trough height in sea state 4 and 6 is only 2\% and 0.16\% less than that in sea state 5, and the deep-to-shallow trough depth ratio changes by less than 4\% both in sea state 4 and 6. 

\begin{figure}
\centering
\includegraphics[width=3.8in,clip=false]{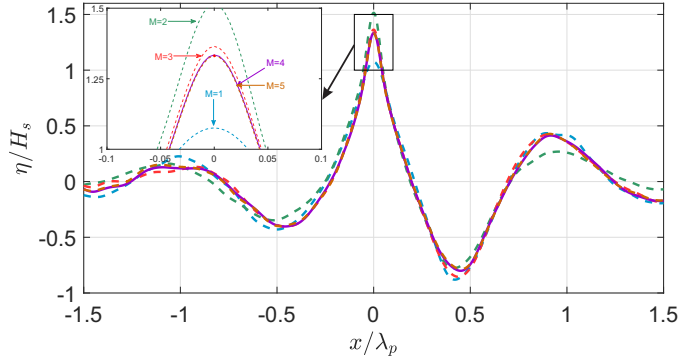}
\vspace{-1.0cm}
\caption{Effect of nonlinearity on the average rogue wave profile in sea state 5 for $M=1-5$. The trough asymmetry is obtained for all $M$ using method ii-, with $\bar{\eta}_D/\bar{\eta}_S$ (deepest to shallowest trough ratio) changing by less than 4\% between $M=1$ and $M=5$. In agreement with previous works \cite{Petrova2006}, the crest height is affected by higher nonlinearities $M=3,~4$ and $5$ (c.f. inset figure). }\label{fig:3} 
\end{figure}

Nonlinear effects are of significant importance in formation and morphology of rogue waves, and as a result here we carefully look at the effect of changing the order of nonlinearity on the rogue wave asymmetry. The results are shown in figure \ref{fig:3} and can be summarized as follows: (1) the rogue wave trough asymmetry is captured for all orders of nonlinearity, with only +4\% (+12\%,+7\% and +0.9\%) discrepancy in $\bar{\eta}_D/\bar{\eta}_S$ for $M$=1 ($M$=2, 3 and 4) compared with $M$=5, (2) the linear model ($M$=1) strongly underestimate the mean rogue wave height and crest height (i.e. -8\% and -19\% compared to $M=5$), (3) the averaged rogue wave profile is overestimated for $M=2$ (+7\% and +14\%) and (4) the results for $M=3,~4$ and $5$ are in quantitative agreement (4.3\% and <1\% discrepancy in rogue wave heights for $M=3,~4$ compared to 5). The convergence of the rogue wave profile with $M\geq 3$ confirms that cubic nonlinearity play a dominant role not only in the generation mechanism and occurrence rate of rogue waves \cite{Benjamin1967,Mori2002a}, but also in shaping the rogue wave crests and troughs. The fact that the rogue-wave trough asymmetry is qualitatively obtained for all $M$ ($M\geq 1$) suggests that the primary mechanism responsible is the dispersive dynamics of ocean waves. This result is in agreement with the fact that the distribution of trough depths around the main crest is mostly random, and that the distribution will be wide spread in all realistic oceanic cases. We would like to note that the effect of the trough asymmetry can also be observed from the relatively small ratio of ${\bar{\eta}_C}/{\bar{\eta}_D}=1.69$ ($M=4$), which we obtain when considering the deep trough $\bar{\eta}_D$ using method ii-. This is much smaller compared with ${\bar{\eta}_C}/{\bar{\eta}_D}=2.21$ if we use method -i (for which $\bar{\eta}_D \approx \bar{\eta}_S$), as well as the value obtained using the third order simulations of MNLS equations, which gives the ratio ${\bar{\eta}_C}/{\bar{\eta}_D}=2.2$ \cite{Socquet-Juglard2005a}.

\begin{figure}
\centering
\includegraphics[width=2.7in,clip=false]{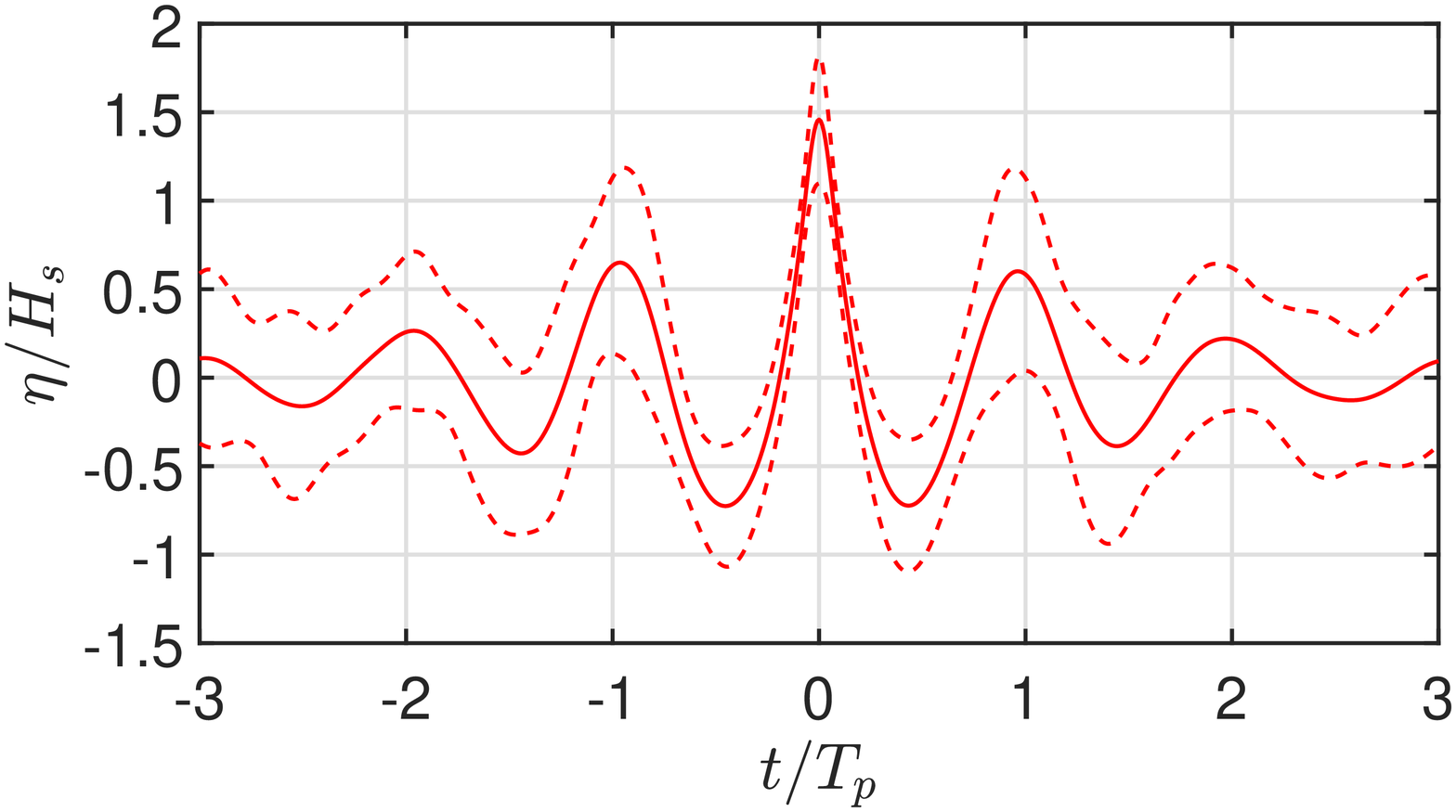}
\put(-165,96){(a)}
\includegraphics[width=2.7in,clip=false]{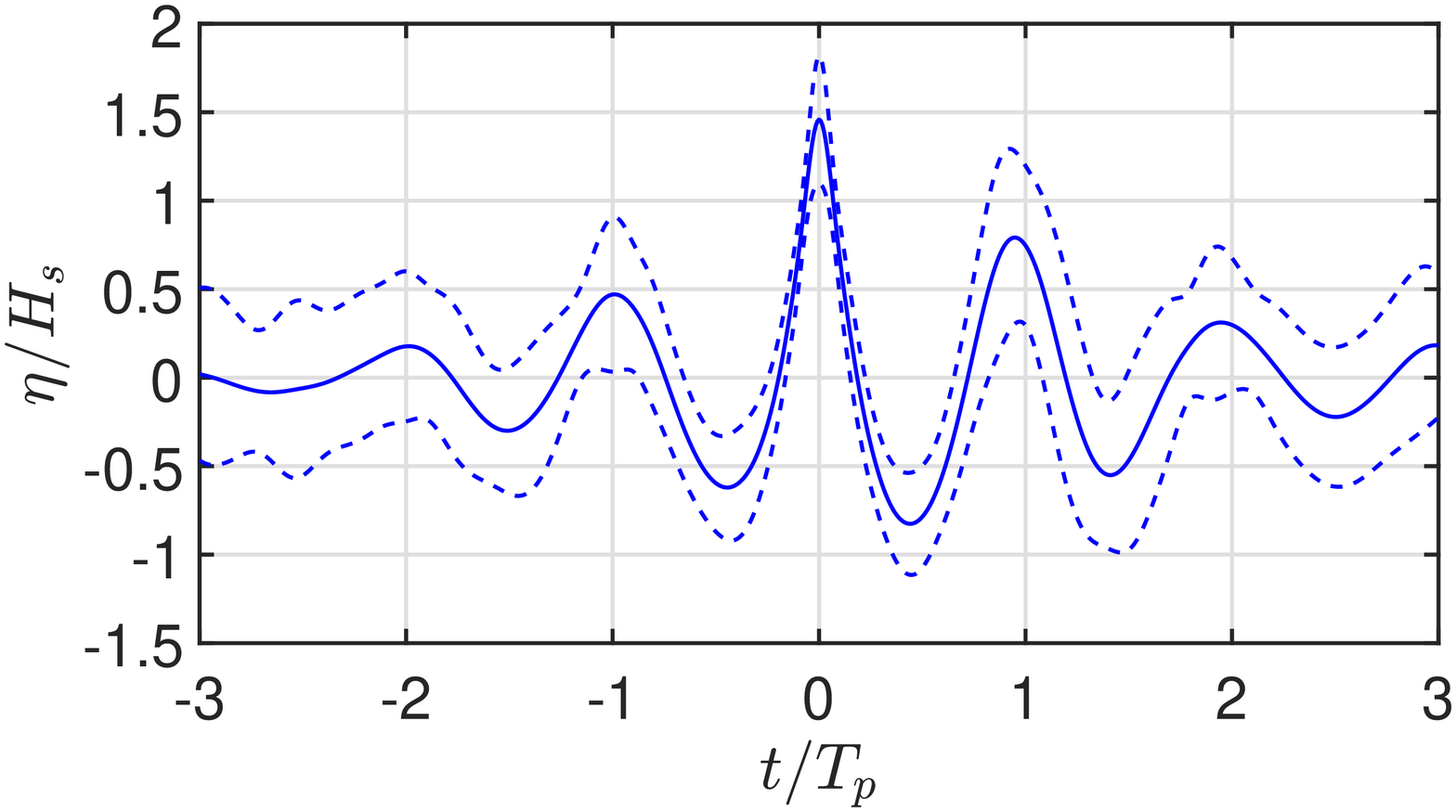}
\put(-165,96){(b)}\\
\includegraphics[width=2.7in,clip=false]{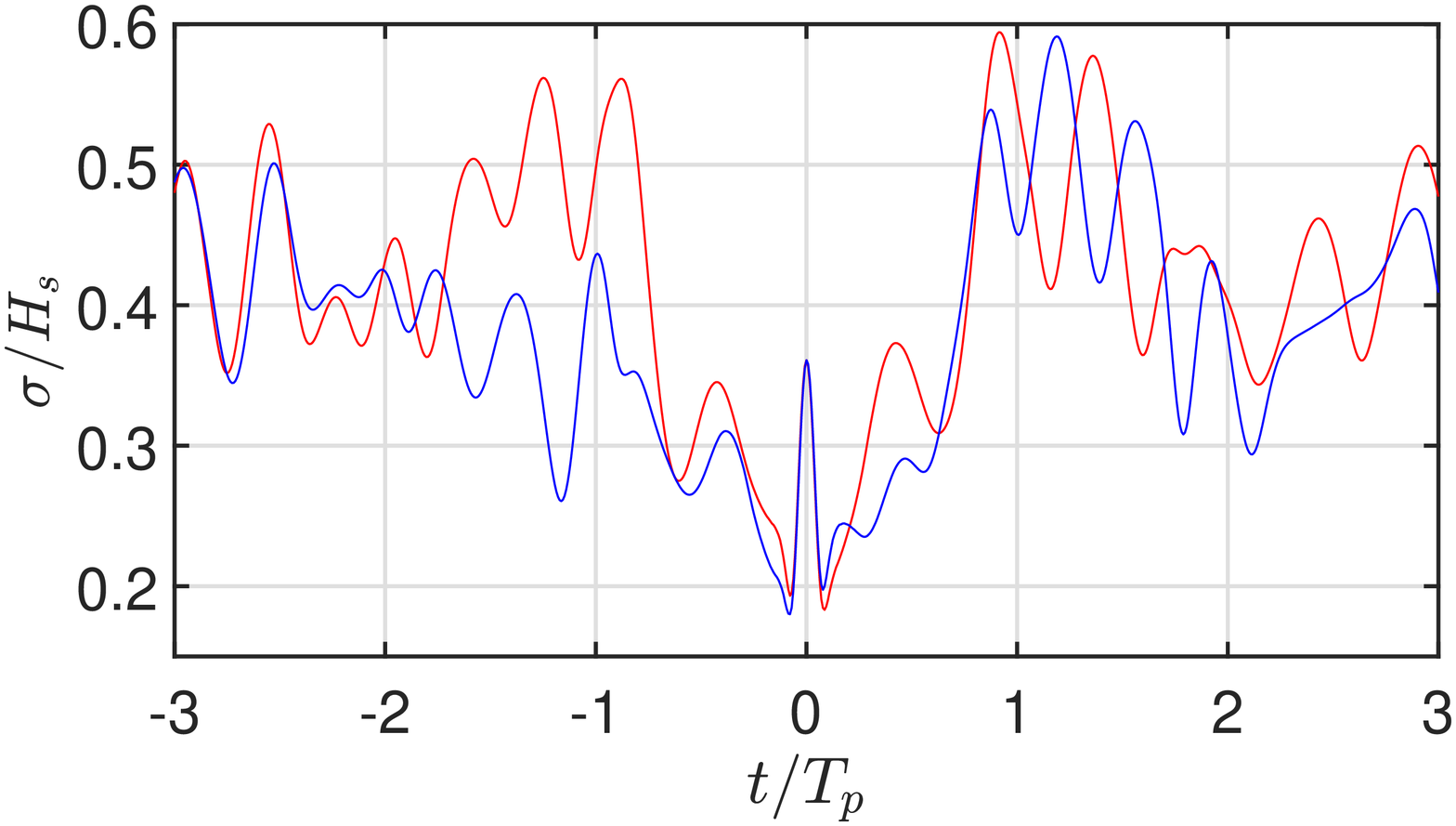}
\put(-165,96){(c)}
\includegraphics[width=2.7in,clip=false]{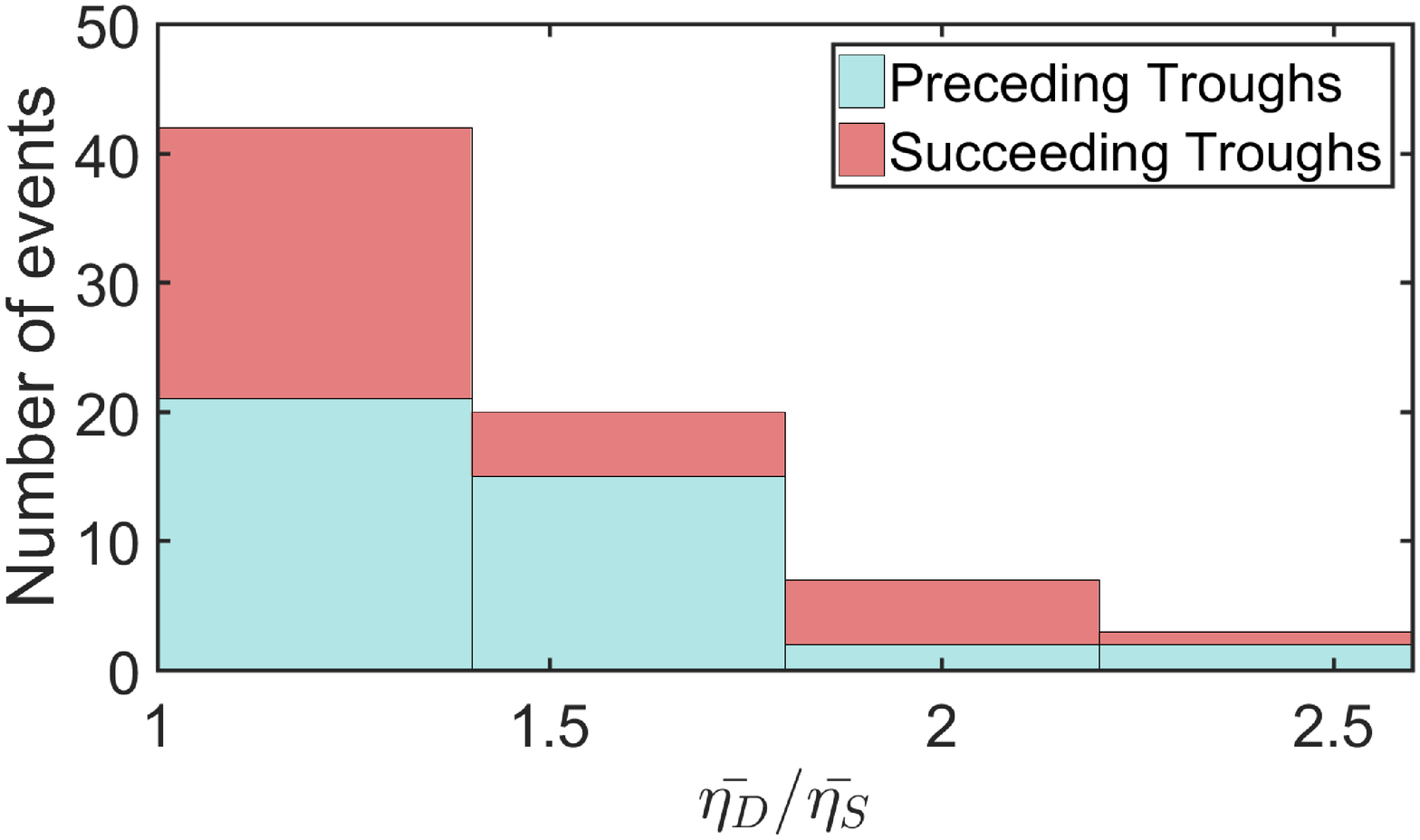}
\put(-165,96){(d)}\\
\caption{Mean temporal rogue wave profile based on (a) method i-, and (b) method ii- for averaging (corresponding to equation \eqref{eq:3} and \eqref{eq:4} with spatial variables substituted with time variables). The trough asymmetry in time is much less pronounced than in space (see figure \ref{fig:2}). The standard deviation for each case is shown in figures (a)-(b) on top of the mean rogue wave profile (dashed lines) and in (c). $\bar{\eta}_D$ to $\bar{\eta}_S$ distribution is shown in the form of two superposed histograms. The bars with blue or red color are the histograms of rogue waves with the deep trough preceding or succeeding the main crest. }\label{fig:5} 
\end{figure}

In figure \ref{fig:5}, we finally show the mean rogue wave profile averaged in time for sea state 5, following methods i- and ii-- but with spatial variables in equations \eqref{eq:3}-\eqref{eq:4} replaced by temporal variables. The temporal wave signal used for averaging is obtained from the simulations using the recorded free-surface elevation at the midpoint between the highest peak and deepest trough of the rogue wave, which is the location that is most likely to capture the high crest and low trough in time. Compared to the mean rogue wave profile in space, the temporal average with flipping displays a relatively small asymmetry. The standard deviation using method ii- is yet still smaller than using method i- (see figure \ref{fig:5}c), confirming that the rogue wave profile in figure \ref{fig:5}b is in better agreement with individual rogue waves than in figure \ref{fig:5}a. Similar to Figure \ref{fig:2}d,  two superposed histograms of the deep-to-shallow trough depth ratio in time domain are shown in \ref{fig:5}d. Again, deeper troughs can travel either preceding or succeeding the main crest.  On average, the deep-to-shallow trough depth ratio $\bar{\eta}_D/\bar{\eta}_S$ is 1.33 for method ii-, and 1.00 for method i-. The largest trough-to-crest height $\bar{\eta}_C+\bar{\eta}_D$ calculated from method ii- is 2.28, which, compared to 2.18 for method i-, is 4.9\% larger.

\section{Conclusion}
The trough asymmetry of rogue waves reported here is a new step toward careful characterization of the most extreme waves in the oceans, which deserve special attention owing to their destructive power. Indeed, the morphology of rogue waves not only sheds light on one of the most important characteristics of the fundamentally-challenging problem of rogue waves, but also is of significant importance in practical applications such as naval architecture and the survivability of man-made offshore structures. We find that the deepest trough extends on average twice as deep as the shallowest one for all sea states 4, 5, and 6 on Douglas scale, which implies an enhancement of the average trough-to-crest height from 1.96 to 2.14 times the significant wave height when considering the deepest trough instead of the trough depth average. The randomness of the trough depths distribution (including preceding/succeeding) is a result of the main mechanism for the asymmetry, which is the linear dispersion of waves. Averages of rogue wave profiles in space thus lose the information on the trough asymmetry when performing direct averaging techniques that remove the preceding/trailing trough depth difference.  

Here we found strong asymmetric troughs of rogue waves especially in spatial domain, which has not been reported from the field  because field observations are often based on buoys' motion in time, for which the signal is much less asymmetric than in space (compare figures \ref{fig:2} and \ref{fig:5}). It may be also because most of our theoretical understanding has come from NLS models, for which solutions exhibit high levels of trough symmetry. NLS-type equations are limited to narrow-band seas, where the dispersion is weak because waves are assumed quasi-monochromatic. This may explain the higher symmetric rogue wave profile of solutions of NLS. It is thus not surprising that solutions to Zakharov equations can show high trough asymmetry.

We expect our results to be relevant to rogue waves in three dimensions, because the mechanism responsible for trough asymmetry is wave dispersion, which dominates wavefield rearrangement in most oceanic conditions. The trough asymmetry of rogue waves in three-dimensions is, however, expected to be more complicated as a result of directional spreading and three-dimensional hydrodynamic instability. For instance, it has long been known that rogue waves can exhibit a fore/aft asymmetry, with the trough following the rogue wave crest bigger than the one preceding (the so-called horseshoe shape), as a result of three-dimensional (class II) modulational instability \cite{Dias1999}. Which one of dispersion or modulational instability is the primary responsible for the formation of rogue waves with asymmetric troughs in the real ocean emerges as a question that would be worth answering.

\section*{JONSWAP spectrum}\label{a1}
The JONSWAP spectrum $\frac{\alpha H_s^2\omega_p^4}{\omega^5}\exp(\beta)\gamma^{\delta}$ used in equation \eqref{init} has constant $\alpha=\frac{1}{16I_{0}(\gamma)}$, where $I_{n}(r)$ is the n-th order moment of the spectrum. The peak enhancement factor $\gamma$ varies from 1 to 9, and with typical value $\gamma=3.3$ we have $I_{0}(3.3)=0.3$. The variables $\beta=-1.25\frac{\omega_{p}}{\omega^4}$ and $\delta=\exp(-\frac{(\omega-\omega_p)^2}{2\omega_p^2\sigma^2})$, where $\sigma$ equals 0.07 and 0.09 for $\omega\leq\omega_{p}$ and $\omega>\omega_{p}$, respectively.

\bibliographystyle{unsrt}
\bibliography{501_JONSWAP_roguewave_prediction,512_profile,513_Rogue_wave_phenomenon,514_Shallow_rogue_wave,515_crossing_seas,516_design,refs}

\end{document}